\documentclass[aps,prb,twocolumn,showpacs]{revtex4}
\usepackage{graphicx}
\begin{document}

\title{Highly anisotropic energy gap in superconducting Ba(Fe$_{0.9}$Co$_{0.1}$)$_{2}$As$_{2}$ from optical conductivity measurements}

\author{T. Fischer} \author{A. V. Pronin}\email{a.pronin@fzd.de} \author{J. Wosnitza}

\address{Dresden High Magnetic Field Laboratory (HLD), FZ Dresden-Rossendorf, 01314 Dresden,
Germany}

\author{K. Iida} \author{F. Kurth} \author{S. Haindl} \author{L.
Schultz} \author{B. Holzapfel}

\address{IFW Dresden, Institute for Metallic Materials, 01171 Dresden,
Germany}

\author{E. Schachinger}

\address{Institute of Theoretical and Computational Physics, Graz University
of Technology, 8010 Graz, Austria}

\date{\today}

\begin{abstract}
We have measured the complex dynamical conductivity, $\sigma =
\sigma_{1} + i\sigma_{2}$, of superconducting
Ba(Fe$_{0.9}$Co$_{0.1}$)$_{2}$As$_{2}$ ($T_{c} = 22$ K) at terahertz
frequencies and temperatures 2 - 30 K. In the frequency dependence
of $\sigma_{1}$ below $T_{c}$, we observe clear signatures of the
superconducting energy gap opening. The temperature dependence of
$\sigma_{1}$ demonstrates a pronounced coherence peak at frequencies
below 15 cm$^{-1}$ (1.8 meV). The temperature dependence of the
penetration depth, calculated from $\sigma_{2}$, shows power-law
behavior at the lowest temperatures. Analysis of the conductivity
data with a two-gap model, gives the smaller isotropic $s$-wave gap
of $\Delta_{A} = 3$ meV, while the larger gap is highly anisotropic
with possible nodes and its rms amplitude is $\Delta_{0} = 8$ meV.
Overall, our results are consistent with a two-band superconductor
with an $s_{\pm}$ gap symmetry.

\end{abstract}


\maketitle

\section{Introduction}

The issue of the symmetry of the superconducting order parameter in
the iron-pnictide superconductors remains unsettled. Shortly after
the discovery of superconductivity in these materials
\cite{kamihara} models, which predict the order parameter to change
its sign on different sheets of the Fermi surface (FS), have been
put forward. \cite{mazin1, kuroki1} In the simplest case of the
$s_{\pm}$-wave symmetry, the superconducting gap $\Delta$ can be
parameterized as $\Delta = \Delta_{0}/2 \cdot (\cos k_{x} + \cos
k_{y})$. Because the FS of the pnictides consists of two distinct
set of sheets, centered at the $\Gamma$ (hole pocket) and M
(electron pocket) points of the Brillouin zone, there are no nodes
in this gap.

Many experiments, including e.g. ARPES, are indeed consistent with
this picture. \cite{christianson, evtushinsky, terashima, williams,
hardy} There is, however, a large body of experimental works, which
cannot be easily explained assuming the nodeless gap scenario.
\cite{martin, muschler, salem, goko} Possibly, the gap function of
the iron pnictides is not universal -- whether the gap is nodeless
or not might depend on the compound and on the doping level. In
order to account for this non-universal behavior, models with
relatively large intraband Coulomb repulsion within the FS pockets
have been proposed, resulting in gap symmetries varying between
extended $s$-wave [$\cos(k_{x}) \cos(k_{y})$], $d_{xy}$ wave, and
$d_{x^{2}-y^{2}}$ wave, all allowing for nodes in the order
parameter. \cite{chubukov, moreo, maier, kuroki2}

Currently, the so-called 122 iron-pnictide family (doped
$A$Fe$_2$As$_2$, $A$ = Ba, Sr, Ca) seems to be best suited for
experimental investigations due to the availability of relatively
large single crystals as well as thin epitaxial films of high
quality. \cite{sefat, hiramatsu, katase, iida, lee} In this article
we focus on Ba(Fe$_{1-x}$Co$_{x}$)$_{2}$As$_{2}$. Thermal
conductivity measurements, performed at temperatures down to 50 mK,
indicate a nodeless in-plane gap in this compound. \cite{tanatar,
reid} The gap is reported to remain nodeless at all doping levels --
from under- to overdoped, but the gap anisotropy increases with
doping. This is consistent with APRES results, which systematically
show a full gap in all pnictide compounds´. \cite{evtushinsky,
terashima, kondo, qian}

Optical spectra of superconductors contain information about the
size and the symmetry of the superconducting gaps. Since the last
two years, a large number of reports on optical properties of
pnictides has been published (see Ref. \onlinecite{hu2} for a
review). The low-frequency (terahertz-far-infrared) optical
properties of the electron-doped 122 pnictides have recently been
investigated by a few groups for doping levels near optimal.
\cite{heumen, kim, gorshunov2, wu, perucchi, nakajima, lobo} There
is, however, neither consensus in these reports about the gap sizes,
nor about the number of gaps (although the majority of reports
agrees on a two-gap picture). Obviously, the analysis of the optical
spectra is challenged by the fact, that the gaps are likely to be
different on different sheets of the FS.

\begin{figure}[]
\centering
\includegraphics[width=\columnwidth,clip]{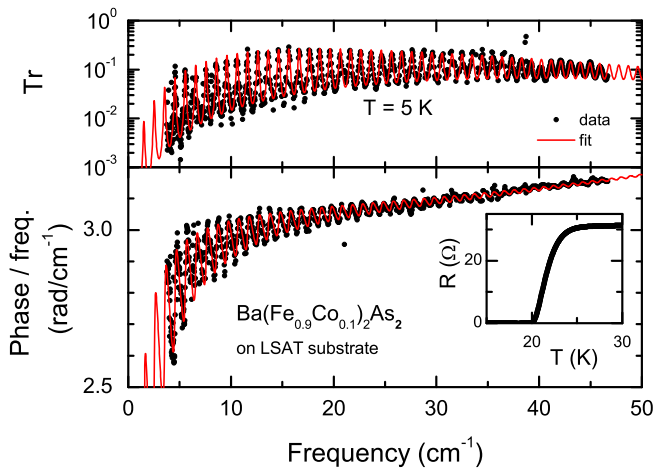}
\vspace{0.2cm} \caption{(Color online) Examples of as-measured
spectra of the transmission coefficient (upper frame) and the phase
shift of the transmitted wave (bottom frame) for a 20 nm
Ba(Fe$_{0.9}$Co$_{0.1}$)$_{2}$As$_{2}$ film on a
(La,Sr)(Al,Ta)O$_{3}$ substrate at $T = 5$ K. The solid lines
represent fit using a two-fluid model. The phase shift spectra are
divided by frequency for better representation. Inset: resistance of
the film around the superconducting transition.} \label{Tr_Pt}
\end{figure}

Our method allows us to perform not only frequency-dependent
measurements, but also measurements of the optical parameters as a
function of temperature at fixed frequencies. This gives advantages
over standard optical studies, since frequency and temperature
dependences of optical parameters can be analyzed simultaneously.

Here, we report on the observation of clear signatures of the
superconducting-gap opening in the terahertz optical conductivity.
Through the temperature-dependent measurements and theoretical
analysis, we show the gap is highly anisotropic with possible nodes
on at least one sheet of the FS.

\section{Experiment}

Films of Ba(Fe$_{0.9}$Co$_{0.1}$)$_{2}$As$_{2}$ have been grown by
pulsed laser deposition on (001)-orientated
(La$_{0.7}$Sr$_{0.3}$)(Al$_{0.65}$Ta$_{0.35}$)O$_{3}$ (LSAT)
substrates, transparent for terahertz radiation. The
Ba(Fe$_{0.9}$Co$_{0.1}$)$_{2}$As$_{2}$ target was ablated with 248
nm KrF radiation under UHV conditions. \cite{iida} The phase purity
was confirmed by x-ray diffraction in Bragg-Brentano geometry. The
$c$ axes of the films was normal to the film surface. The thickness
of the films, used for the measurements, was measured by
ellipsometry. Standard four-probe method has been used to measure
the dc resistivity. We have investigated two
Ba(Fe$_{0.9}$Co$_{0.1}$)$_{2}$As$_{2}$ films with thicknesses of 20
and 100 nm. The results of our investigations for both films are
qualitatively the same. However, the optical transmissivity of the
thinner film is obviously higher, leading to a much better
signal-to-noise ratio. That is why in this paper we present results
for the thinner film only. The resistive onset of the
superconducting transition in the film appeared at 25 K (Fig.
\ref{Tr_Pt}, inset). The substrate was a plane-parallel plate,
approximately $10\times10$ mm in size with thickness of 1.025 mm.

In the frequency range 4 - 47 cm$^{-1}$ (120 - 1400 GHz, 0.5 - 5.8
meV) the measurements have been performed with a spectrometer, which
uses backward-wave oscillators (BWOs) as sources of coherent and
frequency-tunable radiation. \cite{kozlov} A Mach-Zehnder
interferometer arrangement of the spectrometer allows to measure
both the intensity and the phase shift of the wave transmitted
through the Ba(Fe$_{0.9}$Co$_{0.1}$)$_{2}$As$_{2}$ film and the
substrate \cite{gorshunov}. Using the Fresnel optical formulas for
the complex transmission coefficient of the two-layer system, both
components of the complex conductivity ($\sigma_{1} + i\sigma_{2}$)
of the film have been calculated. The optical parameters of the
substrate have been found by measuring a bare substrate without
film. This experimental method has been previously applied to a
large number of different superconductors. \cite{dressel} In
addition to the ``standard" frequency sweeps at fixed temperatures,
we performed temperature sweeps at fixed frequencies in the same way
as it has been done e.g. in Ref. \onlinecite{pronin}. This allows
for a more thorough monitoring of the temperature dependence of
$\sigma_{1}$ and $\sigma_{2}$.

\section{Results}

Figure \ref{Tr_Pt} shows as-measured transmission $Tr(\nu)$ and
phase-shift $\varphi(\nu)$ spectra as a function of frequency, $\nu
= \omega/2 \pi$. The measurements are done with a number of
different BWOs covering the range from 4 to 47 cm$^{-1}$
continuously. Since the major term of the phase shift is
proportional to the frequency of the probing radiation, the
phase-shift spectra are divided by frequency to eliminate the
constant frequency slope. The pronounced fringes in both $Tr(\nu)$
and $\varphi(\nu)$ are due to the multiple interference inside the
substrate, which acts as a Fabry-Perot interferometer. As the
complete Fresnel formulas for a two-layer system are used, these
fringes are automatically taken into account in the calculations of
the complex conductivity.

\begin{figure}[]
\centering
\includegraphics[width=\columnwidth,clip]{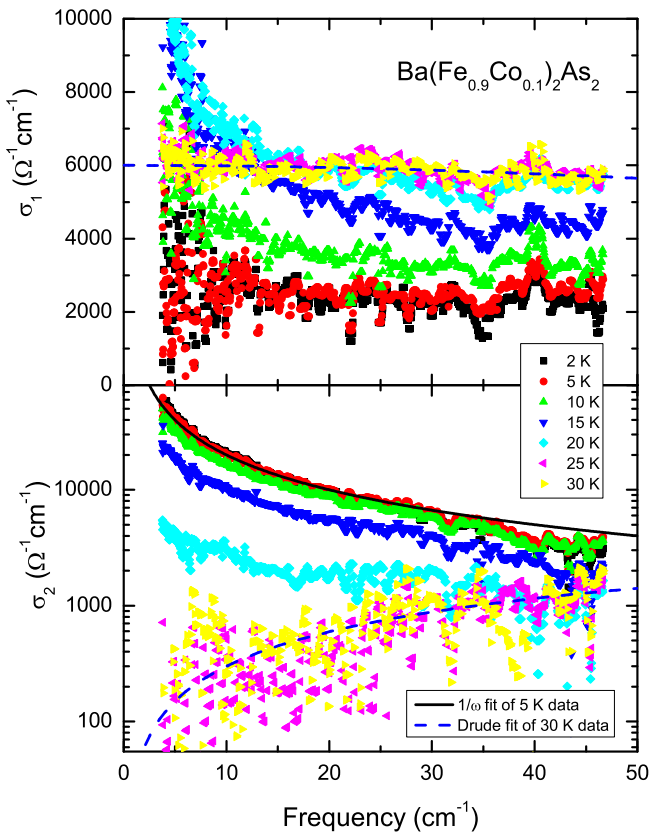}
%\vspace{0.2cm}
\caption{(Color online) Frequency dependence of the real,
$\sigma_{1}$, (upper panel) and the imaginary, $\sigma_{2}$, (bottom
panel) parts of the complex conductivity of
Ba(Fe$_{0.9}$Co$_{0.1}$)$_{2}$As$_{2}$ for selected temperatures
above and below $T_{c}$. The solid line is a fit of the 5 K
$\sigma_{2}$ data with $\sigma_{2} \propto 1/\omega$. The dashed
line is a fit of the 30 K data with a Drude term. Note the
logarithmic Y-scale of the bottom panel.} \label{sg1_sg2}
\end{figure}

Figure \ref{sg1_sg2} shows spectra of the real (upper frame) and
imaginary (bottom frame) part of the complex conductivity of
Ba(Fe$_{0.9}$Co$_{0.1}$)$_{2}$As$_{2}$ at selected temperatures
above and below $T_{c}$.

The normal-state complex conductivity demonstrates typical Drude
behavior, $\sigma(\omega)=\sigma_{dc}/(1-i\omega\tau)$. From
simultaneous fit of $\sigma_{1}(\omega)$ and $\sigma_{2}(\omega)$ at
30 K (dashed line in Fig. \ref{sg1_sg2}), we find $\sigma_{dc}$ =
6000 $\Omega^{-1}$cm$^{-1}$ and the scattering rate $\gamma =
1/2\pi\tau = 200$ cm$^{-1}$ ($\tau = 2.7 \times 10^{-14}$ s), the
former is in agreement with our dc measurements, and the later
coincides with the scattering rate, obtained via $\gamma =
\sigma_{1}(\omega)\omega/\sigma_{2}(\omega)$ from the complex
conductivity spectra ($\gamma$ obtained in this way has, however, a
larger error bar due to the large scattering of data points in
$\sigma_{2}$). From $(\omega_{p}/{2\pi})^{2} = 2 \sigma_{dc}\gamma$,
we estimate the plasma frequency of the Drude component in the
normal state, $\omega_{p}/{2\pi} = 8500$ cm$^{-1}$ (1.05 eV).

When entering the superconducting state the energy gap opens, as
reflected in the lowering of $\sigma_{1}$ below $T_{c}$ (upper frame
of Fig. \ref{sg1_sg2}). For an isotropic $s$-wave superconductor at
$T = 0$ a sharp frequency onset of $\sigma_{1}(\omega)$ is expected
at $2\Delta(0)$. The absence of such a feature in our frequency
window implies that the energy gap $\Delta(0)$ (or one of the gaps
for a multi-band superconductor) must be larger than $47 / 2 = 23.5$
cm$^{-1}$ (2.9 meV). This result is consistent with the majority of
optical measurements in Ba(Fe$_{1-x}$Co$_{x}$)$_{2}$As$_{2}$ -- the
$2\Delta$ feature is usually seen at 50 cm$^{-1}$ or somewhat higher
frequency. \cite{heumen, kim, nakajima, lobo}

The appearance of a $\delta$-function at $\omega = 0$ in
$\sigma_{1}(\omega)$ below $T_{c}$ leads (\textit{via} the
Kramers-Kronig relation) to an $1/\omega$-divergence in
$\sigma_{2}(\omega)$, clearly visible in the bottom frame of Fig.
\ref{sg1_sg2} for $T \leq 20$ K. The strength of this divergence
(the pre-factor) is a direct measure of the weight of the
superconducting condensate. From the $\sigma_{2}(\omega)$ data, one
can directly extract the penetration depth \textit{via}
\begin{equation}
\lambda = c/(4\pi\sigma_{2}\omega)^{1/2}. \label{lam}
\end{equation}
At 2 K, we find $\lambda = 0.45 \pm 0.02$ $\mu$m, which gives for
the plasma frequency of the superconducting condensate, $\omega_{ps}
= c/\lambda$, the value of 3600 cm$^{-1}$ (450 meV), corresponding
to the spectral weight of $(\omega_{ps}/2\pi)^{2} = (1.3 \pm 0.1)
\times 10^{7}$ cm$^{-2}$. These values are within the margins set by
previous reports on terahertz measurements of
Ba(Fe$_{1-x}$Co$_{x}$)$_{2}$As$_{2}$ films. \cite{gorshunov2,
nakamura}

\begin{figure}[b]
\centering
\includegraphics[width=\columnwidth,clip]{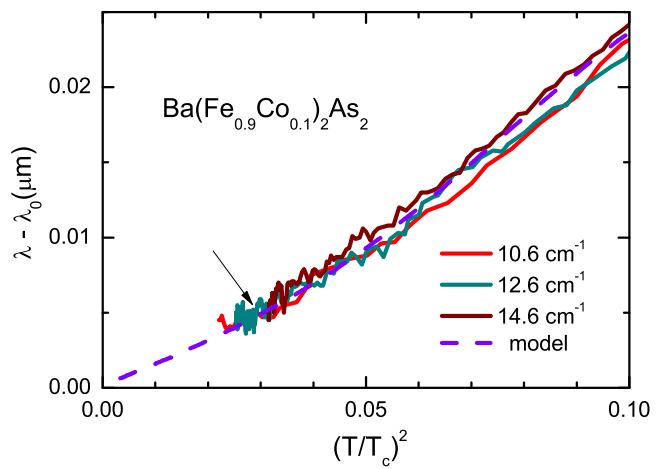}
%\vspace{0.2cm}
\caption{(Color online) Temperature variation of the penetration
depth of Ba(Fe$_{0.9}$Co$_{0.1}$)$_{2}$As$_{2}$ as a function of
$T^{2}$. The solid lines are experimental data. The error bars are
set by the scatting of the data points. The dashed line is
$\lambda(\omega=0)$ obtained with the two-gap model discussed in the
text. The arrow indicates the point where the theoretical line has
been normalized to the experimental value ($(T/T_{c})^{2} = 0.03$).
The mid-point of the transition (22 K) has been taken as $T_{c}$.}
\label{lambda}
\end{figure}

For low enough frequencies and temperatures, where normal carriers
are not affecting $\sigma_{2}$, an analysis of the
temperature-dependent penetration depth is meaningful. We have
extracted $\lambda$ as a function of temperature for a number of
frequencies (Fig. \ref{lambda}). For $T < 0.2 T_{c}$, we observe a
clear power law behavior. As can be seen from this figure, the power
law has an exponent slightly higher than 2, which is in agreements
with microwave measurements. \cite{gordon, prozorov}

\section{Analysis and Discussion}

\begin{figure}[]
\centering
\includegraphics[width=\columnwidth,clip]{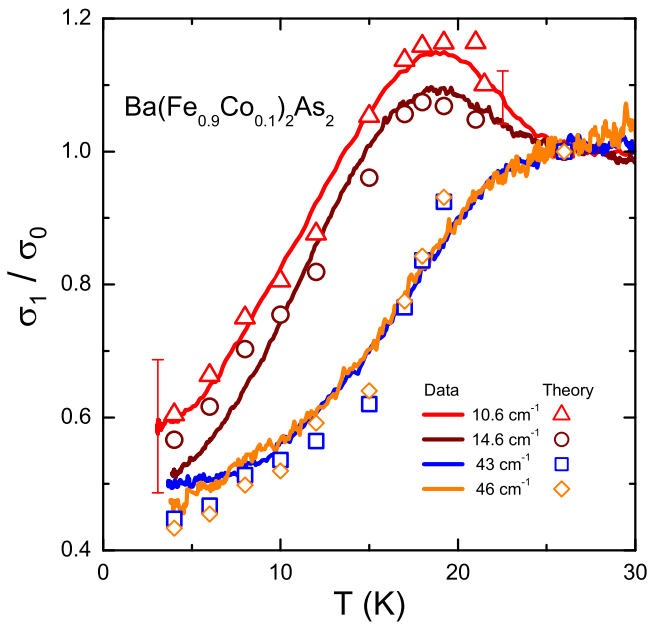}
%\vspace{0.2cm}
\caption{(Color online) Temperature dependence of the real part of
the complex conductivity in Ba(Fe$_{0.9}$Co$_{0.1}$)$_{2}$As$_{2}$.
The values are normalized to the respective normal-state
conductivity at $T = 26$ K ($\sigma_{0}$). Lines - experiment, dots
- calculations with the two-gap model discussed in the text.}
\label{Sg_T}
\end{figure}

Let us now turn to the analysis of the conductivity data. As it can
be seen from the upper panel of Fig. \ref{sg1_sg2}, at frequencies
below 15 cm$^{-1}$, the temperature dependence of $\sigma_{1}$ below
$T_{c}$ is non-monotonic. Upon entering the superconducting state,
$\sigma_{1}$ first rises, and then drops. In order to monitor this
temperature dependence in more detail, we performed temperature
sweeps at fixed frequencies (Fig. \ref{Sg_T}). At low frequencies, a
prominent coherence peak appears in $\sigma_{1}$. Multi-gap effects
and gap symmetry are known to have crucial influence on the
coherence peak appearance. \cite{dolgov, schachinger}

The analysis of the data is performed in the spirit of Ref.
\onlinecite{schachinger}. We use an $s+d$-wave gap \cite{schurrer}
as a model for the extended $s$-wave, which is one part of the
$s_{\pm}$ model. We can do this because optics cannot resolve the
phase difference, thus for the optics the $s_{\pm}$ model just
consist of one isotropic $s$-wave gap ($\Delta_{A}$, hole pocket)
and a second extended $s$-wave gap, which, in our case, is modeled
using an $s+d$-wave symmetric gap when referred to the M-point. This
model is also described in greater detail in Ref.
\onlinecite{carbotte}.

The gap on the electron pocket is parameterized as $\Delta_{B} =
\Delta_{s} + \Delta_{d} \sqrt{2} \cos(2\theta)$ (here $\Delta_{s}$
is the $s$-wave component, $\Delta_{d}$ is the amplitude of the
$d$-wave component, and $\theta$ is the polar angle on the FS). An
anisotropy parameter $\alpha$ has been introduced in such a way that
$\Delta_{s} = \alpha\Delta_{0}$, $\Delta_{d} = \sqrt{1-\alpha^{2}}
\Delta_{0}$, and $\Delta_{0} =
\sqrt{\langle\Delta_{B}^{2}(\theta)\rangle_{\theta}}$
($\langle...\rangle_{\theta}$ is the FS average). In the clean
limit, this gap has nodes, if $\alpha \leq \sqrt{2/3}$.

For convenience, one can introduce the parameter $X = \alpha /
(\alpha + \sqrt{1-\alpha^{2}}) \times 100\%$, which gives the
percentage of the $s$-wave component in the $s+d$-wave admixture.
The temperature dependence of the superconducting gaps has been
modeled by the standard mean-field BCS temperature dependence with
$T_{c} = 22$ K (the mid-point of the transition). The complex
conductivity is a sum of two weighted components, corresponding to
the two FS pockets: $\sigma(\omega,T) = w_{1}\sigma^{(1)}(\omega,T)
+ w_{2}\sigma^{(2)}(\omega,T)$, and $w_{1}+ w_{2} = 1$.

In our analysis we concentrated on the 10.6 cm$^{-1}$ data set
following the procedure outlined by Schachinger and Carbotte.
\cite{schachinger} The size of the two gaps, the $s$-wave
contribution ($X$) to the anisotropic gap, and the weights $w_1$ and
$w_2$ were adjusted in order to reproduce the temperature dependence
of the normalized conductivity $\sigma_1/\sigma_0$ as good as
possible. We obtained the best fit with $\Delta_A = 3$ meV = 24
cm$^{-1}$, $\Delta_{0} = 8$ meV = 64 cm$^{-1}$, and $X = 33.5 \%$,
corresponding to $\alpha = 0.45$ and indicating the existence of
nodes. The weights of the FS pockets were found to be $w_{1} = 20
\%$ for the hole and $w_{2} = 80 \%$ for the electron pocket, the
normal-state elastic scattering rate being equal to 22 meV (180
cm$^{-1}$). \cite{wu1}

The model calculations for the other three frequencies (14.6, 43,
and 46 cm$^{-1}$) are found to follow the experimental data without
any further adjustment of the primary parameters. The theory
predicts, assuming an instantaneous transition, that the 43 and 46
cm$^{-1}$ data sets should also show a coherence peak at $T \approx
21.5$ K. This temperature, however, is already well within the
superconducting transition of the film and thus, such peaks cannot
be observed experimentally. What is left as a very convincing result
is the correct reproduction of the coherence peak for 10.6 and 14.6
cm$^{-1}$ and a very good agreement with the data measured at 43 and
46 cm$^{-1}$ for $T \leq 20$ K.

We calculated, moreover, using the model discussed previously the
low temperature variation of the penetration depth in the zero
frequency limit. \cite{modre} The result is presented in Fig.
\ref{lambda} as the dashed line. At $(T/T_{c})^{2} < 0.1$, the
agreement between the model and the experimental data is perfect.
\cite{lambda}

Another explanation for the low-temperature $T^2$ dependence of the
penetration depth was given by Vorontsov \textit{et al.}
\cite{vorontsov} These authors studied the superconducting $s_\pm$
symmetry state and concluded that the strong interband interaction
(scattering) is strongly pair-breaking and that the superconductor
is driven into a gapless state. For this state the authors prove
that the penetration depth will be $\propto T^2$ at low
temperatures. A similar result was reported by Gordon \textit{et
al.} \cite{gordon2} who studied a highly anisotropic single $s$-wave
superconductor with a high concentration of pair-breaking impurities
which will also drive the superconductor into a gapless state. They
argued that because of the assumed high anisotropy this model must
not necessarily be in contradiction to the thermal conductivity
results of Tanatar \textit{et al.} \cite{tanatar}

Authors of recent optical measurements of
Ba(Fe$_{1-x}$Co$_{x}$)$_{2}$As$_{2}$ argue for interpretation of the
conductivity data in terms of an $s_\pm$ state with strong
pair-breaking scattering. \cite{lobo, aguilar} Nevertheless, it is
necessary to point out that Nicol and Carbotte \cite{nicol} studied
the optical properties of a classical $s$-wave superconductor with
magnetic impurities. They come to the conclusion that in the gapless
state (strong pair-braking) a coherence peak becomes strongly
suppressed which is in clear contradiction to our results presented
in Fig. \ref{Sg_T}. Likely, as it has been proposed in Ref.
\onlinecite{lobo}, both effects, the large gap anisotropy and the
pair-braking scattering have to be taken into account in order to
describe the optical conductivity in the pnictides. No detailed
model for such a calculation has been developed yet. The analysis
performed in Ref. \onlinecite{lobo}, however, suggests that only a
minor part of $\sigma_{1}$ in the superconducting state below the
gap can be due to the pair-breaking effects, while the major
contribution is due to the nodes or at least strong gap anisotropy.

Our observation of a superconducting gap with nodes is, at a first
glance, at odds with the results of Tanatar \textit{et al.}
\cite{tanatar} However, the $s+d$-wave model used by us to describe
the anisotropic superconducting gap around the $M$-point of the
Brillouin zone provides a mechanism called ``lifting of the nodes",
first discussed by Mishra \textit{et al.} \cite{mishra} and later on
studied explicitly in Ref. \onlinecite{carbotte}. The model allows a
lifting of the nodes in the superconducting gap given a high enough
$s$-wave contribution to the $s+d$-wave admixture and additional
elastic scattering. This mechanism is also temperature dependent and
it is more likely to lift existing nodes the lower the temperature.
Thus, it is possible within this model that at very low
temperatures, like the ones used by Tanatar \textit{et al.},
\cite{tanatar} a small spectral gap can develop in the real part of
the optical conductivity. Increasing the temperature will smear out
this small gap and $\sigma_1$ will show nodal behavior at higher
temperatures. Most recent thermal conductivity measurements give
evidence of a small gap (almost nodes) in regions of the FS which
contribute significantly to in-plane conduction. \cite{reid}

\section{Conclusions}

We found the superconducting energy gap in
Ba(Fe$_{0.9}$Co$_{0.1}$)$_{2}$As$_{2}$ to be highly anisotropic.
This anisotropy is responsible for the broad and pronounced
coherence peak observed in $\sigma_{1}$ which is in clear
contradiction  to a gapless superconducting state. A two-gap model
-- isotropic $s$-wave gap for the hole pocket and an $s+d$-wave gap
for the electron pocket of the FS (Ref. \onlinecite{schachinger}) --
has been applied to mimic the $s_{\pm}$ gap symmetry. The model
provides a very good description of the $\sigma_{1}(T)$ curves
measured at different frequencies. In the framework of this model,
we have found the value of 3 meV for the isotropic gap and an rms
amplitude of 8 meV for the second gap which is also found to be
nodal. It was possible using this model to reproduce consistently
not only the energy and temperature dependence of $\sigma_1$ in the
terahertz region but also the low temperature variation of the
penetration depth.

\section{Acknowledgements}

We would like to thank Jules P. Carbotte and Ilya Eremin for useful
discussions and Mykola Vinnichenko for the film-thickness
measurements. Part of this work was supported by EuroMagNET II (EU
contract No. 228043).


\begin{thebibliography}{99}
\bibitem{lambda} Because our measurements are performed at relatively high
frequencies, the contribution of the normal carriers to $\sigma_{2}$
(and, consequently, to $\lambda$, Eq. \ref{lam}) becomes not
negligible at higher temperatures, causing the experimetal data to
deviate from the model curve calculated for $\omega=0$.
\end{thebibliography}
\end{document}